\pdfoutput=1
\documentclass[runningheads]{article}
\usepackage[utf8]{inputenc}
\usepackage{graphicx}
\usepackage{cite}
\usepackage{hyperref}
\title{Cloudless Friend-to-Friend Networking for Smartphones\thanks{This paper is a revised and extended version of Cloudless Wide Area Friend-to-Friend Networking Middleware for Smartphones \cite{icete18} ICETE 2018 conference paper. The extended version includes the novel concept of reconnectable channels (section \ref{reconnectablechannels}), which has never been published before, including the accompanying figures. Other changes include title and keywords, entirely rewritten abstract, introduction, sections 2 and 3, and conclusion. The other sections have also been revised. We have replaced the many footnotes with URLs with inline parentheses. Some webpages are now references instead of footnotes. Statistics about Pv6 adoption and Tor relays have been updated to more recent numbers. A citation about smartphone usage has been removed from the introduction and replaced by another.}}
\date{}
\author{Jo Inge Arnes and Randi Karlsen \\
University of Tromsø - The Arctic University of Norway, Tromsø, Norway}

\begin{document}

\noindent\makebox[\linewidth]{\rule{\paperwidth}{0.4pt}}

\Large{The final authenticated publication is available online at \url{https://doi.org/10.1007/978-3-030-34866-3_10}}

\noindent\makebox[\linewidth]{\rule{\paperwidth}{0.4pt}}

{\let\newpage\relax\maketitle}

\begin{abstract}
Using smartphones for peer-to-peer communication over the Internet is difficult without the aid of centralized services. These centralized services, which usually reside in the cloud, are necessary for brokering communication between peers, and all communication must pass through them. A reason for this is that smartphones lack publicly reachable IP addresses. Also, because people carry their smartphones with them, smartphones will often disconnect from one network and connect to another. Smartphones can also go offline. Additionally, a network of trusted peers (or friends) requires a directory of known peers, authentication mechanisms, and secure communication channels. In this paper, we propose a peer-to-peer middleware that provides these features without the need for centralized services.

Mobile peer-to-peer \and Friend-to-friend networking \and Unreachable IP addresses \and Location transparency.
\end{abstract}
\section{Introduction}
Around 2.5 billion people in the world use smartphones \cite{statista}. People use smartphone apps for a wide range of online services that are important to them, such as apps for news, banking, education, career, and health. It is also common to use smartphones for social networking, where the users communicate with each other and share pictures and videos.

Smartphone apps are typically backed by services in the cloud. The apps connect to clouds running within large data centers, and the cloud services handle most of the apps' data storage and processing needs. By using a centralized cloud service, it is also easier to share data between smartphones. For example, when someone shares a picture via Snapchat (https://www.snapchat.com), the picture is uploaded to Snapchat's cloud. A cloud service then sends notifications to the user's friends. The friends open the picture in their Snapchat app, which downloads the picture from Snapchat's cloud. The pattern is typical for how smartphone apps communicate and share data.

The clouds thus represent centralized hubs for communication and data management, which can pose various privacy problems. One problem is that the impact of data breaches can be massive. Another problem is that many companies actively gather, analyze, and sell user information on a large scale \cite{datatilsynetUnhappy, datatilsynetRace}. Removing the dependence on centralized services may be a step towards alleviating these issues. Modern smartphones are also computers with processing, memory, and storage capabilities comparable to regular desktop PCs less than a decade ago. With the increasing popularity of smartphones compared to PCs \cite{Myers16}, it may be sensible to make more use of the smartphone's local hardware as an alternative to cloud computing. For example, smartphones can participate as nodes in a distributed system that combine their storage and processing capabilities. Such a system should allow smartphones to communicate directly instead of via centralized services. Cloud-based solutions are, in essence, client-server architectures. The alternative is to use peer-to-peer communication. However, when communicating over the Internet, smartphones rarely have reachable addresses, and they often change networks. It is challenging to find ways to connect smartphones over the Internet and to keep track of network locations, which is why apps depend on clouds or other remote application services for orchestrating communication.

We present a novel approach to smartphone peer-to-peer over the Internet, which aims to solve the problem of frequent network changes and lack of publicly reachable addresses. We introduce Swirlwave, a middleware that does not rely on clouds and client-server architectures – we term this \emph{cloudless}. Swirlwave handles peer-to-peer communication well in experiments. Smartphones are directly reachable, and addresses are automatically updated when peers change networks. We also suggest an approach enabling continued communication when a smartphone disconnects from a network and connects to another.      

We first describe related work. Next, the architecture and communication methods are explained. We then describe the middleware, followed by experiments and results. Finally, we discuss the findings before concluding the paper.

\section{Related Work}
Turtle \cite{Popescu06} is a theoretical friend-to-friend architecture for safely sharing sensitive data, where the system floods search queries throughout the network. Turtle builds on trusted relationships between people when defining friend-to-friend networks, which is also true for  Swirlwave. However, Turtle is theoretical and does not address how to connect devices. Swirlwave is a middleware that connects the smartphones as peers over the Internet, and it is not restricted to a particular use case.

Orbot (https://guardianproject.info/apps/orbot) is an official app for communicating over the Tor network. Swirlwave uses Tor to make smartphones directly connectable, and the implementation of Swirlwave uses some of the same underlying libraries as Orbot. Unlike Swirlwave, Orbot does not solve what happens when the smartphone changes to another network. It also has none of Swirlwave's friend-to-friend networking features. 

Thali (http://thaliproject.org) is an experimental platform for peer-to-peer web solutions. The project is open-source and sponsored by Microsoft. The aim is to enable the creation of apps that take advantage of the smartphone's resources and give users better control over their data. This is comparable to Swirlwave's objectives. Thali, however, gave up the idea of peer-to-peer over the Internet using Onion services \cite{toronionservices}, then called hidden services. The project concluded that the Onion service protocol only is usable for stationary devices \cite{thalihiddenservices}. They chose instead to focus on Bluetooth Low Energy (BLE), Bluetooth, and Wi-Fi Direct. None of these are for wide-area communication. Swirlwave supports wide-area communication over the Internet.

\section{Mobile Peer-to-Peer Communication Without Public IP Address}

In this section, We explain why non-public IP addresses is a challenge, our solution, and the architecture of the system.

\subsection{Unreachable Addresses}
\label{Addresses}

When a smartphone connects to the Internet, it is usually behind network address translation (NAT). The smartphone is, in reality, connected to a local area network (LAN) and communicates with the Internet through a router. Only the router's address is visible from the Internet, e.g., a server will only see the router's address when it is contacted by a smartphone. The server replies to the router, which routes the traffic to the smartphone on the LAN \cite{Comer14}. The IP addresses within the LAN are not valid outside. Additionally, local IP addresses are commonly assigned dynamically by a DHCP-server, so the smartphone's address can change each time it reconnects to the LAN. The consequence of DHCP is that the addresses are unpredictable inside the LAN. More importantly, the consequence of NAT is that others cannot reach the smartphone from the Internet, which prevents peer-to-peer communication. This is true both for Wi-Fi and cellular data.

\subsection{Architecture}
\label{Swirlwave-ch}

Swirlwave belongs to a family of peer-to-peer architectures called friend-to-friend networking \cite{bricklin2000friend}. Friend-to-friend networks are unstructured and private. Peers only connect directly to already known peers (friends) \cite{rogers2007disappear}. The friendships are mutual (commutative), but not transitive, which means that a friend of a friend is not automatically accepted.

\begin{figure}
	\includegraphics[width = 5.5cm]{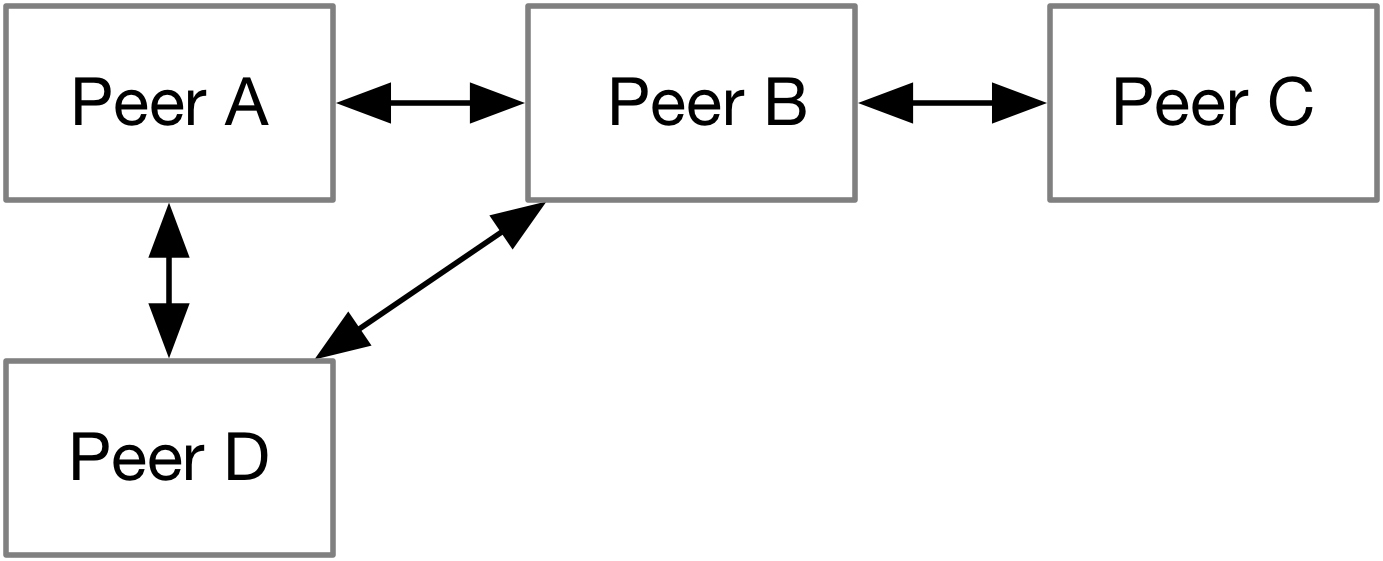}
	\caption{Friend-to-friend network \cite{icete18}}
	\label{F2F-fig}
\end{figure}

Figure \ref{F2F-fig} illustrates that A, B, and D are friends that all know each other. The same does not apply to C, which is a friend of B but does not know anyone else. Peer C can only contact B directly. Nevertheless, C can reach others indirectly if B relays requests to its friends. A and D can relay the request to other friends, and so on, thus enabling far-reaching queries.

Friend-to-friend networks can be used for distributed systems where the included nodes should not be known outside the system. For example, a company can connect its smartphones as part of a closed system. Friend-to-friend networks can also be used to define private social networks. The peers in the network can act as clients and servers at the same time. They can provide services that other peers can use. At the same time, they can be clients that consume services provided by others. 

Swirlwave is a middleware that hides the complexity of friend-to-friend networking from the apps that use it. The apps do not know the location of the other peers or that Tor is used as an underlying protocol. Instead, they connect to localhost with ordinary TCP sockets when they wish to communicate with remote peers. The middleware automatically routes traffic between smartphones. 

\begin{figure}
	\includegraphics[width = 7.0cm]{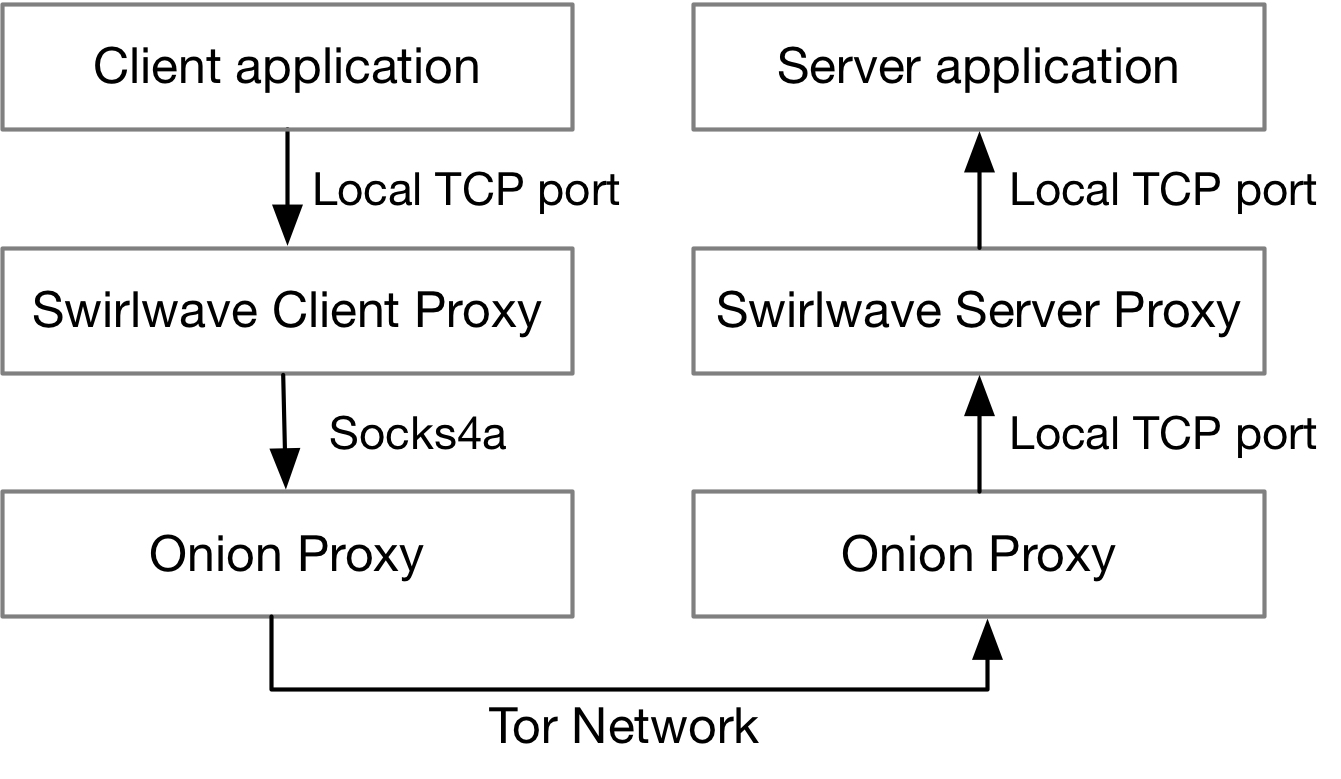}
	\caption{Proxying from client to server \cite{icete18}}
	\label{Proxys-fig}
\end{figure}

Swirlwave defines two proxies for handling the routing: The client-proxy, and the server-proxy. Both run locally on the smartphone. Figure \ref{Proxys-fig} illustrates two peers, where the left peer acts as a client, and the right peer acts as a server. The app layer connects to the locally running client proxy via TCP. It can do this because the client proxy listens to a range of ports. Each port numbers is associated with a specific friend and service. The client proxy then uses the SOCKS4a \cite{Lee12} protocol to connect to a friend through a locally running Tor onion service. 

\subsection{Tor and Tor Onion Services}
Tor (https://www.torproject.org) was made for anonymity, while Swirlwave uses it for connectivity purposes. 

Tor conceals online activity by routing encrypted traffic through layers of dedicated onion routers, often compared to the layers of an onion. The Tor network is public and has more than 6,500 running volunteer relays, per July 2019 (https://metrics.torproject.org). All traffic passes through at least three onion routers before reaching the destination — the onion routers at each end of a connection act as proxies for entering the network. The onion proxy on the client side has a directory of available onion routers. It picks out the routers that it wants to use, and Tor builds a circuit. Each router in the circuit only knows its successor and predecessor.     The onion proxy at the end of the connection sends traffic to the final destination, which is an ordinary server that is unaware of Tor. The protocol conceals the client's identity from the server, not the other way around.

If we wish to hide the server's location from clients, we can use the Tor Onion Service protocol (previously called hidden services). In this protocol, the server registers itself with the Tor network to obtain an onion address. Clients can reach the server through the Tor network by using this address. The protocol aims to ensure that the server's location remains unknown to the clients.

Swirlwave utilizes an unintended consequence of the onion service protocol, which is NAT traversal. When a server registers as an onion service and obtains an onion address, it must actively connect to the Tor network. To be available from the network, it keeps this connection open. If the server is behind NAT, the protocol still works because NAT only hinders connections coming in from the Internet, not in the opposite direction \cite{Comer14}. The consequence is that we can obtain an address that can reach devices behind NAT. Tor also has the advantage of being a public overlay network that is open to anyone.

Swirlwave uses onion services to reach smartphones over the Internet. It runs an onion proxy locally at each smartphone. However, Tor does not solve all the challenges. Smartphones frequently change networks as people carries them around in their daily life. Onion services do not work well in this scenario. When a smartphone changes its location, it can theoretically continue to use the same onion address. The problem is that other smartphones that have already built circuits to it will continue to send traffic to the old location. Also, there is no protocol transparency. Apps must use the SOCKS4a protocol and know how about onion services to use them. The Swirlwave middleware provides this missing location and protocol transparency.

\section{Swirlwave}
\label{Contact}

Swirlwave is used to build friend-to-friend networks. Each peer keeps a directory of friends. The middleware automatically manages addresses changes and updates the directory. Swirlwave uses onion addresses to reach peers without publicly visible IP addresses. Anonymity is central in Tor, but Swirlwave requires authentication of friends. Authentication is thus part of Swirlwave.

\subsection{Contacts}

Keeping track of peer addresses is one of Swirlwave's central features. It achieves this without external directory services or central points. Each peer in Swirlwave keeps a local directory of known peers. New contacts are added out-of-band, for example, by using near-field communication (NFC).

An entry in the contact list contains data that is needed to communicate with a peer. The entry can also include information about services offered by the peer. Entries contain the peer ID, onion address, services provided by the peer, phone number (used for the SMS fallback protocol) and its public-key. See Table \ref{Table:ContactList} for details.
\begin{table}
	\caption{Information in a contact list record \cite{icete18}}
	\label{Table:ContactList}
	\begin{tabular}{|l|l|}
		
		\hline
		Field name & Description \\
		\hline
		
		Name & A human-readable name of the friend \\
		
		Peer ID & An ID that is unique across all installations \\
		
		Address & The friend`s onion-address \\
		
		Address Version & Each time a peer changes its address, it will increment the address version number. \\
			
		Secondary Address & The phone number used when sending SMS-messages to the peer \\
		
		Public-key & The public-key from the friend's asymmetric keys \\
		
		Online Status & Offline if last attempt to reach the friend was unsuccessful, otherwise online \\
		
		Last Contact Time & The last time contact was made with the peer \\
	
		Known Friends & A list of peer IDs for mutual friends \\
		
		Capabilities & A list of capabilities supported by this friend. Such as available services and protocol UUIDs \\
		
		Awaiting Answer & A flag indicating if an answer from the SMS fallback protocol is pending  \\
		
		\hline
	\end{tabular}	
\end{table}

A server and its clients must use the same protocol to conduct meaningful communication. Swirlwave lets the application layer decide which protocol to use. This flexibility is possible by using universally unique identifiers (UUID) \cite{leach2005} as identifiers for protocols. The identifier identifies a communication protocol that the server and client must recognize in order to communicate. Swirlwave uses the identifier to match clients and servers. 

For example, to send a message to a friend, a user selects the friend from the Swirlwave contact list. Based on the protocol UUIDs registered for the friend, Swirlwave presents a list of supported protocols. If the user has a client-app that understands the protocol, Swirlwave detects it by matching the identifiers of the local app with the identifier of the friend's service.

\subsection{Authentication and Confidentiality}

Each peer has a key-pair for public-key encryption \cite{Goodrich14}. The key-pair is used for authentication purposes. Also, Swirlwave can use it for ensuring confidentiality, integrity, and non-repudiation of data when communicating over other channels than Tor. When using onion services, the traffic is end-to-end encrypted, so this already provides the needed communication confidentiality. 

Anonymity is usually an essential feature of Tor. The onion proxy on the server side will not know the origin of incoming connections. In our approach, this anonymity hinders the identification of incoming requests from friends. Swirlwave solves this by providing an authentication mechanism that validates the incoming connections. This functionality is part of the Swirlwave proxies and based on public-key cryptography. 

When a client-peer wants to establish a new connection, the client-side proxy sends a system message to the server-peer. The client has encrypted the message with its private key. The server-side proxy decrypts the message with the client's public key to validate the identity of the sender. If the server-peer cannot validate the identity, it refuses the connection.

\subsection{Establishing Connections}

\begin{figure}
	\includegraphics[width = 5.0cm]{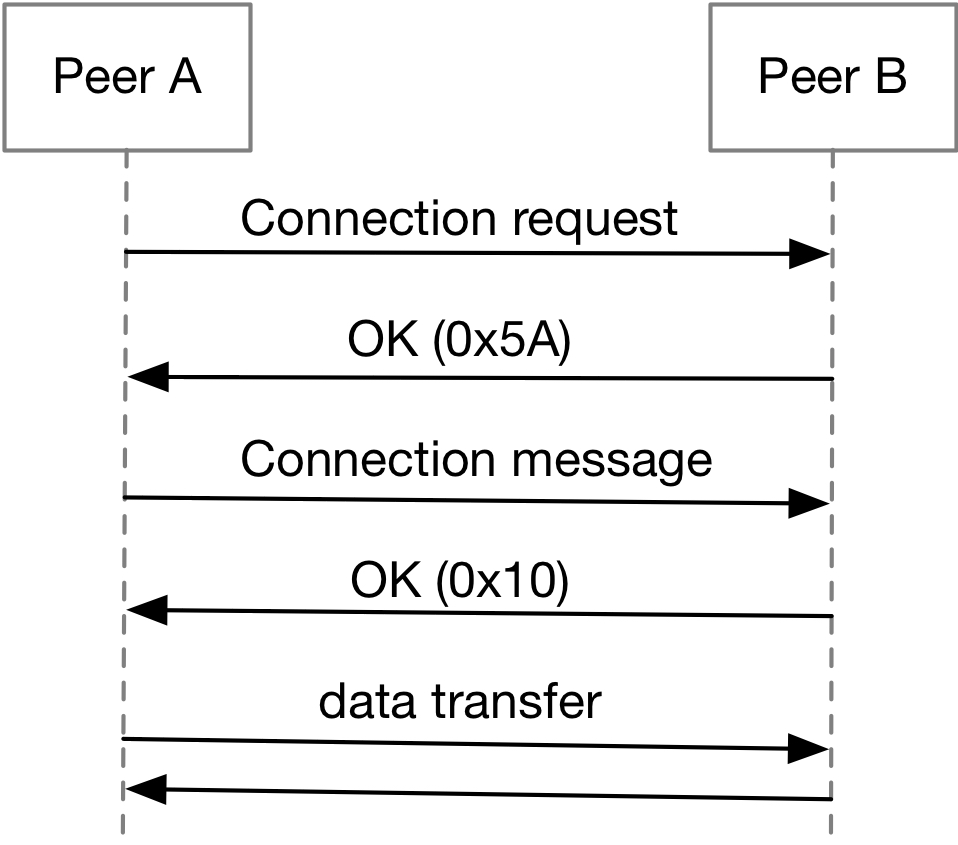}
	\caption{Establishing connection \cite{icete18}}
	\label{Connection}
\end{figure}

To establish a connection, a request is sent from a Swirlwave client proxy via the onion proxy.  The message header contains (among others) the friend's onion-address.  If the onion proxy returns a positive response code (0x5A) telling that it successfully connected to the remote onion service, the client proxy also receives a four-byte number.  This number is later returned to the server as part of the connection message.  If the server proxy accepts the connection (after assessing the connection message) it responds with a success code (0x10).  The client proxy then begins reading and writing bytes between the incoming socket from the application-layer client and the outbound onion proxy socket. Figure \ref{Connection} illustrates this communication.

Everything in a connection message, except the client ID, is encrypted with the client's private-key.  The server proxy looks up the peer ID in the contact list and rejects the connection if it does not recognize the peer. If the peer ID is in the contact list, its public-key is used to decrypt the message.  If the message successfully is decrypted, and the returned number equals the one that the server earlier sent to the client, the connection is approved.  The connection message also specifies whether the connection will be used for for transmitting system messages or application-layer data.  For an application-layer connection, the server proxy will use an identifier found in the destination field, match it to a local service endpoint, and establish a connection to the application-layer service.  For a system message, instead the content of the message field is dispatched to an internal module that manages system messages. See Table \ref{Table:Connection} for a list of the information included in the connection message.
\begin{table}
	\caption{Connection message information \cite{icete18}}
	\label{Table:Connection}
	\begin{tabular}{|l|l|}
		
		\hline
		Field name & Description \\
		\hline

		Sender ID & The peer ID of the client \\
		
		Random Number & A random number initially generated and sent by the server proxy \\
		
		Message Type & Whether this is a system message or an application-layer connection \\
		
		Destination & An identifier of a capability representing a service that the client wishes to consume. \\
		& This will only be set for application-layer connections. \\
		
		System Message & A system message that will be dispatched to a module that handles system messages. \\
		& This will only be set for system message types. \\
		
		\hline
		
	\end{tabular}
\end{table}

When the onion proxy fails to connect to the remote onion service, the client proxy marks the peer as being offline. It then begins the process of getting an updated address to the peer, either by asking a mutual friend, or by using an SMS fallback protocol which contacts the peer directly. The client proxy will not attempt to establish new connections to the friend until an updated address is obtained. The friend will then be marked as online again.

\subsection{Address Changes}

When a smartphone moves from one network to another, for example, from cellular data to Wi-Fi, its access point is no longer the same. The IP address will most likely be different, and the route to the device will definitely be different.

Such address changes must be announced to friends. A device that has been offline, or has changed its location, will contact its friends as soon as it is online again. The new address is passed with a version number. The version number increases whenever a peer changes address, and it is used to determine which is the most recent address when comparing across peers.

After a peer has been offline, it is not unlikely that some friends have changed their addresses. The peer will not have received the updated address, and will not be able to reach them. The previously offline peer can either ask mutual friends about the friend's new address or contact the friend directly via a fallback protocol using SMS.

\begin{figure}
	\includegraphics[width = 6.0cm]{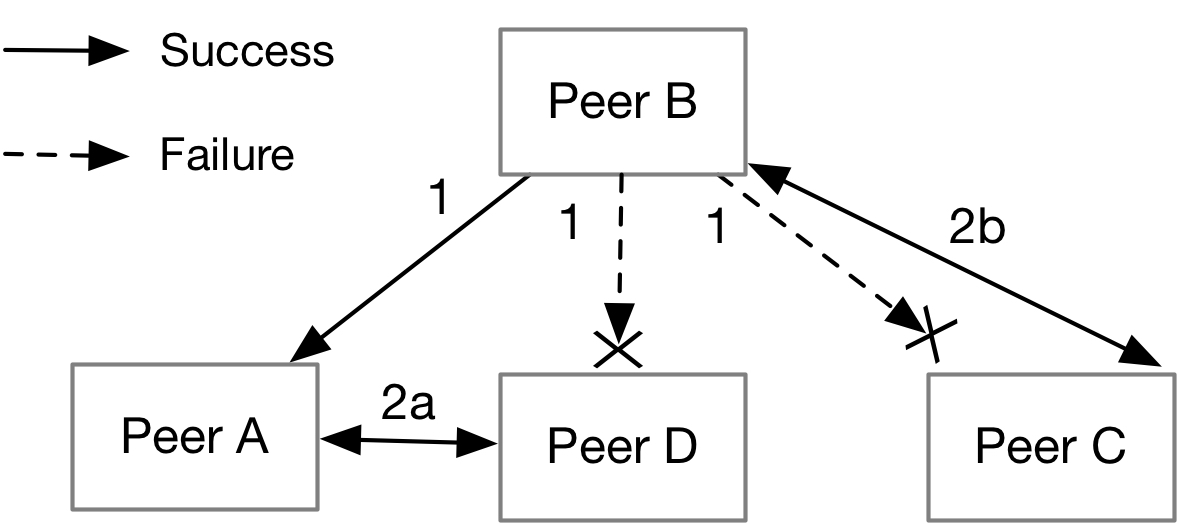}
	\caption{Peer B changes address \cite{icete18}}
	\label{Change address-fig}
\end{figure}

Assume that peer B, in Figure \ref{Change address-fig}, has changed its address. It sends a system message (marked 1) including the new address to three friends. Peer A successfully receives the address and updates its contact list. The other peers cannot be reached. When C and D later tries to contact B, they discover that B is unreachable. They will then try to obtain an updated address. Peer A is a mutual friend of D and B. D can ask A for B's address (marked 2a). Peer C, on the other hand, does not have any other friends to ask. Instead, it uses an SMS to ask B directly for its address (marked 2b). Phone numbers can be seen as stable addresses that will work even when a peer has changed its network location.  

Swirlwave uses onion addresses to connect to peers. Routing is still dependent on IP addresses under the hood, just as with all Internet traffic. When a peer changes its network, the IP and route to it change as well. While it is possible to reuse an onion address at the new location, there is no support for letting clients refresh the route to the onion service. Swirlwave thus monitors network changes and registers a new onion service when the smartphone connects to a new network. An exception is if a network and access point address is recognized from earlier. Then the onion service and address from last time can be reused.

\subsection{SMS Fallback Protocol}

An SMS fallback protocol is used to request new addresses from friends when they cannot found at their previously known onion address. When the client proxy of peer $C$ discovers that it cannot connect to a friend, $B$, and there are no other friends to ask for the address, it will send a data SMS to $B$. Unlike a text SMS, a data SMS will not be visible to the user. It will instead be received by the Swirlwave middleware running on the smartphone.

The fallback protocol begins by $C$ sending $B$ an SMS that includes $C$'s address and a secret one-time code. $C$ has encrypted the message with its private key. The one-time code has several purposes; it enables duplicate message detection, and it is a message-ID and anti-forgery token that later will be returned to $C$.

$B$ looks up $C$'s phone number in its contact directory, to confirm that the SMS is from a known friend. $B$ decrypts the message and updates the contact list with $C$'s address. If Swirlwave is running and connected, an answer immediately is sent to $C$ over the Internet including $B$'s current address. Otherwise, $C$ will get an answer as soon as $B$ reconnects. In the answer, $B$ sends its new address together with the one-time code that $C$ sent earlier. $B$ has encrypted the message with its private key. $C$ updates the contact list with $B$'s new address when it receives the message.

\subsection{Reconnectable Channels}
\label{reconnectablechannels}

So far, we have described how peers automatically announce address changes and how they keep track of each other. However, when a peer disconnects from one network location to connect to another, it will necessarily break any open connections to its peers. The peer will not be available for others until it has finished the process of changing its network location. This frequent breaking of connections is undesirable from an application perspective, and it can affect the user experience negatively. The underlying middleware should instead hide the instabilities from the application, whenever possible.

In this section, we propose the concept of reconnectable channels. The aim is that if peers A and B are communicating over a connection, and peer A changes its network, then the applications on A and B can continue unaffected for a while. Within this time frame, A has the opportunity to reconnect to B and resume communication from where it got lost. If successful, the applications will be unaware of the temporary disconnect.

The enabling factor for implementing reconnectable channels is the use of client and server proxies, which already provide location and protocol transparency. When two peers communicate, all traffic passes through the local proxies on each side. On both sides, each proxy keeps two sockets: One socket connects the proxy to a locally running application, the other socket connects the proxy to the external network. The proxy sits in between the two sockets and transfers data back and forth, from one to the other.

\begin{figure}
\includegraphics[width=4.0cm,height=4.0cm]{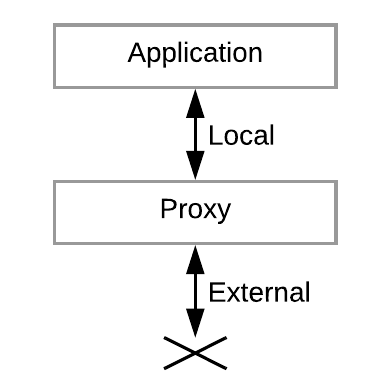}
\caption{The connection between the proxy and the external peer is closed, but the connection between the application and the proxy is still open.} 
\label{Local-proxy-external-connection-fig}
\end{figure}

When a peer disconnects from the network, the connections between the proxies and the external network inevitably become closed on both sides of the ongoing communication. However, the situation is very different for the sockets that connect the proxies and local applications. These connections are purely local and are unaffected by the status of the external network. Thus, they are still available to the applications. Figure \ref{Local-proxy-external-connection-fig} shows that the application connection is still open after the external connection has closed.

For a limited time, a local proxy can hide that the connection to the external network has closed from an application. It does this by keeping the connection to the application open and simulate a low transfer rate. The proxy continues to read and write data at a rate just high enough to keep the connection alive by preventing a timeout. The application can receive and send data as before.

\begin{figure}
\includegraphics[width=8.0cm,height=4.0cm]{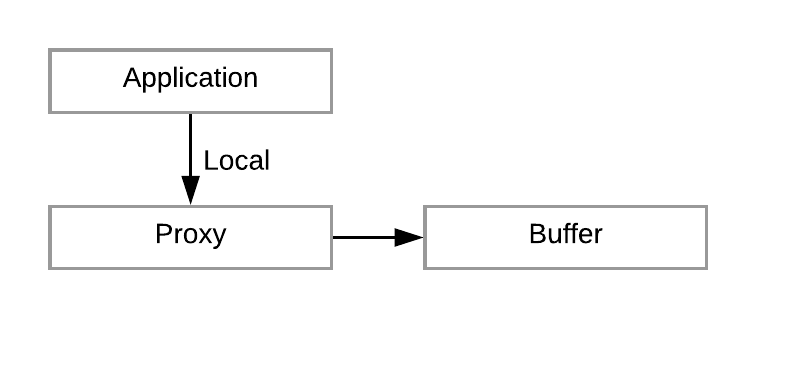}
\caption{The external connection is temporarily unavailable, and the proxy has started using buffers instead.} 
\label{proxy-buffering-data-fig}
\end{figure}

The proxy continues to receive data from the application, but it has nowhere to send the data because the intended recipient is temporarily disconnected. Instead, the proxy stores the data in a buffer. If the external recipient successfully reconnects within a reasonable time, it will send the buffered data to it. This is illustrated in Figure \ref{proxy-buffering-data-fig}.

Broken TCP-connections are only detected when trying to send data, not during reading \cite{cleary}. The sender expects an acknowledgment from the other end of the connection that the data was received. Reading data from a connection in TCP is a passive act and does require an acknowledgment. An application will not detect a broken connection through reading operations. The phenomenon is known as a half-open connection. There is no need for the proxy to send data to the application to hide that the external connection is down.

After a device has finished changing its network location, it will attempt to restore the previously ongoing communication with its peers. The proxy will open sockets to the peers, and send reconnection messages requesting to reestablish and continue the communication. On success, the proxies on both sides of the communication will start sending their buffered data over the external network, now at high speed. The local proxy does not read at maximum speed from the application socket until it has finished sending the buffered data to the external peer.

Another challenge that arises is that data can become lost on its way from the source to the destination when a connection closes abruptly. There will be a discrepancy between what the source (A) has sent and what the destination (B) has received. After reestablishing the communication between A and B, the data transfer should continue from the last data that B received. It should not continue from the last data that A sent. 

Consequently, B must tell A where to resume the data transfer. The sender (A) must always remember the most recent data that it has sent to B. The proxies achieve this through the use of buffers. A proxy will gradually fill a buffer with data as it has sent, and when the buffer is full, the proxy will begin filling it up from the beginning. The proxy does not clear the buffer before starting at the beginning again, so the buffer constitutes a sliding window over the data that A already has sent. On the other side, B will maintain a counter for how much data it has received, which B sends to A on a reconnect. The modulo of the buffer size can be used to calculate the index of the last received data and if the data is still available. In this way, it is possible to resume the communication between A and B without any loss of data due to a disconnect.

Lastly, each peer can have multiple connections to many other peers simultaneously. To properly reconnect, each connection has an associated identifier. If A wants to reconnect to B, it has to send a message including this identifier. B will then look up information locally to verify that A indeed was the other side of the communication before the connection broke.

In this section, we have introduced the concept of reconnectable channels, which seamlessly enables applications to continue communicating with peers even when connections break due to network location changes on either side.

\section{Experiments}
\label{Experiments}

We conducted several experiments with the middleware. Testing included the functionality of the system, as well as location transparent communication, performance measurements, startup time, connection establishment time, throughput, transmission time, and latency. In this section, we describe the results. 

\subsection{Peer-to-Peer Communication}

We have tested using Swirlwave for establishing connections over the network, connecting to peers, and transferring data. In the experiments, we also tested how the system handles location changes. 

In one experiment, two smartphones connected as peers. A webcam app ran on one smartphone, and the other smartphone showed the live stream in a browser. Swirlwave made this possible while both smartphones were connected to 4G from different providers, without the usually required streaming via clouds or similar centralized services. We also demonstrated that smartphones were able to change between 4G and Wi-Fi during streaming. In that case, the smartphones update each other's onion addresses, and the streaming continues. 

The browser on the client smartphone connects to a port on localhost. Thus, from the browser's point of view, the webcam seems to be on the local smartphone. However, Swirlwave manages the current onion address to the peer and routes the traffic to the correct smartphone via Tor. This means the browser can continue to use the localhost-address and can be kept unaware of network changes. This is a demonstration of location transparency and streaming from anywhere without being connected to Wi-Fi.

\subsection{Performance}

We compared with two alternative configurations when evaluating the performance; one where we used Tor directly without Swirlwave, and one where we used a plain Internet connection. 

We used two smartphones during the experiments. A Huawei P9 Lite was used as a client. It was connected to the Internet via cellular data (4G).  A Samsung Galaxy Note 4 was used as a server. It was connected to the Internet via Wi-Fi. The experiments were carried out within a short time-frame. 

Orbot, which is the official version of the Tor onion routing service on Android, was used to test Tor without Swirlwave. Orbot can be used to connect the smartphones, but it has no mechanism for changing addresses when a smartphone changes location. It is still enough for conducting the experiments.

We used an Internet subscription with a static, public IP address for the experiments, which enabled us to manually assign a publicly visible IP to the smartphone that acted as a server. A wireless router was manually configured to forward from a specific port on that smartphone. The server smartphone could then be contacted directly from the Internet. The limitation of this approach, compared to Swirlwave, is that the server smartphone cannot be reached as soon as it leaves the manually configured Wi-Fi. Also, the smartphone cannot take the role as a server when connected to cellular data.

\subsubsection{Starting Onion Proxy}

The experiment measured how long it took from the onion proxy was started until the onion service was registered and ready for use. The difference between starting a new onion service and reusing an existing one was compared.

The implementation of Swirlwave can reuse an already registered onion service when it reconnects to a network location recognized from before. As seen in Table \ref{Table:Start}, there is a difference in onion proxy start-up times between registering a new onion service and reconnecting to one that is already registered (which skips the registration process). Registering a new onion service took around twice as long. The start-up time also varied more when registering new onion services compared to reusing an address. When comparing the 90th percentiles, the start-up time when reusing an address was about five times faster than when registering a new onion address.
\begin{table}
	\caption{\small{Onion proxy start-up times \cite{icete18}}}
	\label{Table:Start}
	\begin{tabular}{|l|l|l|l|}
		
		\hline
		& Median & $90^{th}$ Percentile & Num. Trials \\		
		\hline
		
		New onion service & 18.080s & 43.941s & 10 \\
		
		Reused onion service & 8.401s & 8.855s & 10 \\
		
		\hline
		
	\end{tabular}
\end{table}

\subsubsection{Establishing Connections}

How long it took to establish a connection between client and server for the three different cases:

\begin{itemize}
	\item Connecting via Swirlwave. This includes the time it takes to authenticate the client.
	\item Both client and server use Orbot.
	\item Connecting directly over the Internet.
\end{itemize}
\begin{table}
	\caption{Times for establishing connection \cite{icete18}}
	\label{Table:Connections}
		\begin{tabular}{|l|l|l|l|}
			
			\hline
			& Median & $95^{th}$ Percentile & Num. Trials \\
			\hline
			
			Connecting via Swirlwave & 1.829s & 3.557s & 100 \\
			
			Onion service w/Orbot & 1.384s & 2.931s & 100 \\
			
			Directly over Internet & 1.827s & 3.597s & 100 \\
			
			\hline
			
		\end{tabular}
\end{table}

Our experiments showed that the time to establish connections was nearly identical when connecting via Swirlwave and directly via Internet. This was surprising because the connections in case of Swirlwave must be made through the Tor network. Also, the connection times for Swirlwave include client authentication. Connecting via Tor using Orbot, which does not include any authentication, was faster than connecting directly via the Internet. This may suggest that the difference between connecting via Tor and via the Internet roughly equals the time Swirlwave uses to authenticate the client.

\subsubsection{Throughput}

Throughput is the rate of successful data delivery over a communication channel \cite{Forouzan13}. Given that a connection was established and the client was authenticated, we measured how long it took from the client started reading the first byte until 12.5MB had been read. The rate was then calculated.
\begin{table}
	\caption{Throughput \cite{icete18}}
	\label{Table:Throughput }
	\begin{tabular}{|l|l|l|l|}
		
		\hline
		& Median & $95^{th}$ Percentile & Num. Trials \\
		\hline
		
		Swirlwave & 2.510Mbps & 1.380Mbps & 74 \\
		
		Onion service w/Orbot & 1.950Mbps & 0.910Mbps & 100 \\
		
		Directly over Internet & 18.58Mbps & 11.95Mbps & 100 \\
		
		\hline
		
	\end{tabular}
\end{table}

The throughput was lower when routing via Tor than directly over the Internet. This was true for both Swirlwave and Orbot. The throughput for Swirlwave was higher than Orbot. Transmitting directly over the Internet without Tor was 7.4 times faster than Swirlwave, and 9.5 times faster than Orbot.

\subsubsection{Transmission Time}

Transmission time is the time it takes from the first bit until the last bit of a message is sent from a node. Transmission time depends on message size and bandwidth \cite{Forouzan13}, as shown in equation \ref{eq:Transmission}.
\begin{equation} \label{eq:Transmission}
Transmission time = MessageSize / Bandwidth		
\end{equation}
When we estimated the transmission time, the median throughput was used in place of the bandwidth, and the message size was set to 8 bits.
\begin{table}\footnotesize
	\caption{Transmission times \cite{icete18}}
	\label{Table:Transmission}
	\begin{tabular}{|l|l|}
		
		\hline
		& Transmission Time 1 Byte (8 bits) \\
		\hline

		Swirlwave & $3.200\times10^{-6}$s (3.200$\mu$s) \\
		
		Onion Service w/Orbot & $4.103\times10^{-6}$s (4.103$\mu$s) \\
		
		Directly over Internet & $4.306\times10^{-7}$s (0.4306$\mu$s) \\
		
		\hline
		
	\end{tabular}
\end{table}

The results show that transmission time for all  alternatives were very low, with the Internet-connection having the lowest result, followed by Swirlwave and Orbot.

\subsubsection{Latency}

Network latency specifies how long it takes for a bit of data to travel across the network from one node to another \cite{Forouzan13}. Latency depends on several components, as shown in equation \ref{eq:Latency}.
\begin{equation} \label{eq:Latency}
	Latency = Propagation Time + Transmission Time + 
	Queuing Time + Processing Time
\end{equation}
We first measured round-trip time (RTT), which is the time it takes from the client sends a byte until it receives a response byte from the server. This had the advantage that start and end times could be measured at the same smartphone. RTT is described in equation \ref{eq:RTT}
\begin{equation} \label{eq:RTT}
RTT = 2 \times Latency + Processing Delay
\end{equation}
The extra processing delay represents the time from the byte is read by the server until it sends a response byte to the client.
\begin{table}
	\caption{Round-Trip Times \cite{icete18}}
	\label{Table:Round}
	\begin{tabular}{|l|l|l|l|}
		
		\hline
		& Median & $95^{th}$ Percentile & Num. Trials \\
		
		\hline
		Swirlwave & 0.637s & 0.816s & 100 \\
		
		\hline
		Onion service w/Orbot & 0.639s & 1.554s & 100 \\
		
		\hline
		Directly over Internet & 0.106s & 1.039s & 100 \\
		
		\hline	
	\end{tabular}
\end{table}

We estimated latency based on the RTT measures, using the simplified calculation shown in equation \ref{eq:LatencySimpl}.
\begin{equation} \label{eq:LatencySimpl}
    Latency = RTT / 2
\end{equation}
\begin{table}
	\caption{Latency \cite{icete18}}
	\label{Table:Latency}
	\begin{tabular}{|l|l|l|}
		
		\hline
		& RTT median & Latency \\
		\hline

		Swirlwave & 0.637s & 0.3185s \\
		
		Onion service w/Orbot & 0.639s & 0.3195s \\
		
		Directly over Internet & 0.106s & 0.053s \\
		
		\hline
		
	\end{tabular}
\end{table}

Latencies were almost similar for Swirlwave and Orbot, about three tens of a second. When transmitting directly over the Internet, the latencies were approximately six times lower.

From Table \ref{Table:Transmission} we see that transmission times were negligible when considering latency. Latency in Table \ref{Table:Latency} therefore depend on propagation time, plus the time used for processing and queuing in the nodes.

\section{Discussions}

To connect to hosts behind NAT from the Internet, we need techniques for NAT Traversal \cite{Hu05}. These techniques circumvent the problems associated with address translations and private IP addresses in different ways. 

We considered several approaches to NAT traversal before deciding to use Tor instead, including Virtual Private Network (VPN) \cite{Comer14}, UDP hole punching \cite{Hu05}, and SSH (https://www.ssh.com/ssh). All of the alternatives had drawbacks that made them less desirable for use with Swirlwave. For example, setting up VPN servers requires public IPs and much manual work for configuration and management. Also, for clients to work as servents (as both client and server), they need reserved IP addresses or other mechanisms for locating peers. Regarding UDP hole punching, the technique only supports UDP. It also needs a server middlebox to establish peer-to-peer communications. We considered SSH, but it requires a server with a public IP address.

We also chose to use SMS as part of a fallback alternative to find addresses of friends for which the address was no longer known. We considered gossiping, hand-offs, and other techniques \cite{Tanenbaum14}, but they are all more complicated, less secure, less reliable, and require more than two peers. We found that the SMS fallback protocol has several advantages. SMS is available on all smartphones, telephone numbers are a stable address, and the protocol works even when only two peers exist in the friend-to-friend network. Also, it does not require extra hardware, servers, or software to work. 

The functionality and performance of Swirlwave were measured experimentally. We find that Swirlwave handles peer-to-peer communication between smartphones well. Smartphones can act as both client and servers, and continued communication is also possible when the devices move between networks. From the experiments, we found that the downside is the lower throughput caused by Tor. The extra round trips and processing involved in authenticating the client does not seem to affect the performance considerably. The same is true for the processing done by the Swirlwave proxies. Experiments show that it takes time to establish a connection between peers. Connections should thus be kept open if possible, instead of closing after each shorter session has finished. 

In the paper, $IP$ means IPv4 \cite{postel1981rfc}. IPv4 is by far the most widespread version as of today. However, with the more recent IPv6 \cite{deering1998internet} version of IP, each device will be given a public address and NAT will in theory no longer be needed. However, a form of NAT, or something similar, that hides the real IPv6-address will probably be standard for privacy and security reasons. An addition to the IPv6 protocol, called mobile IPv6, is designed to let devices keep their address even when changing networks \cite{perkins2011mobility}. In this scenario, Swirlwave would not need Tor for connectivity, but would rather build on IPv6. The adoption of IPv6 is still low in most countries. Per June 2019, it is about 26.9\% in the U.K., 15.1\% in Norway and 46.5\% in the U.S. \cite{akamaiip6}. We believe, therefore, that Swirlwave (or similar types of middleware) will continue to be useful in the future.

\section{Conclusion}

Wide area peer-to-peer for smartphones usually relies on cloud services as brokers of communication. These solutions thus have more in common with client-server architectures than peer-to-peer. We have presented a middleware that aims to remove the need for centralized services, which is not trivial. We have explained the many challenges regarding peer-to-peer for smartphones and provided solutions for them integrated into the middleware. The middleware hides the details from the application layer, which can remain unaware of aspects such as lack of publicly reachable IP addresses, location changes, authentication, and security. Through the concept of reconnectable channels, the applications can also continue to operate during disconnects in conjunction with network changes.


%
%
%
\bibliographystyle{splncs04}
\bibliography{md}
\end{document}